\documentclass[12pt]{article}
\usepackage{graphicx,amsmath}
\usepackage{units}

\newlength{\dinwidth}
\newlength{\dinmargin}
\renewcommand{\vec}[1]{\boldsymbol{#1}}

\def\lapproxeq{\lower .7ex\hbox{$\;\stackrel{\textstyle                                                    
<}{\sim}\;$}}                                                    
\def\gapproxeq{\lower .7ex\hbox{$\;\stackrel{\textstyle                                                    
>}{\sim}\;$}}                                                    
\def\be{\begin{equation}}                                                    
\def\ee{\end{equation}}                                                    
\def\bea{\begin{eqnarray}}                                                    
\def\eea{\end{eqnarray}}
\def\b{\vec{b}}
     
\def\q{\vec{q}}

\def\GeV{\rm GeV}

\def\sh{\hat s}
\def\sh2{{\hat s}^2}
\def\sd{$\sigma^{\rm D}_{{\rm low}M}~$}
\def\se{$\sigma_{\rm el}~$}
\begin{document}
                                                    
\titlepage                                                    
\begin{flushright}                                                    
IPPP/13/37  \\
DCPT/13/74 \\                                                    
\today \\                                                    
\end{flushright} 
\vspace*{0.5cm}
\begin{center}                                                    
{\Large \bf Diffraction at the LHC}\\

\vspace*{1cm}
                                                   
V.A. Khoze$^{a,b}$, A.D. Martin$^a$ and M.G. Ryskin$^{a,b}$ \\                                                    
                                                   
\vspace*{0.5cm}                                                    
$^a$ Institute for Particle Physics Phenomenology, University of Durham, Durham, DH1 3LE \\                                                   
$^b$ Petersburg Nuclear Physics Institute, NRC Kurchatov Institute, Gatchina, St.~Petersburg, 188300, Russia

\vspace*{1cm}                                                    
                                                    
\begin{abstract}                                                    

We show that the diffractive $pp$ (and $p\bar{p})$ data (on $\sigma_{\rm tot}$, $d$\se$/dt$, proton dissociation into low-mass systems, \sd, and high-mass dissociation, $d\sigma /d(\Delta\eta)$) in a wide energy range from CERN-ISR to LHC energies, may be described in a two-channel eikonal model with only one `effective' pomeron.  By allowing the pomeron coupling to the diffractive eigenstates to depend on the collider energy (as is expected theoretically) we are able to explain the low value of \sd measured at the LHC.  We calculate the survival probability, $S^2$, of a rapidity gap to survive `soft rescattering'. We emphasize that the values found for $S^2$ are particularly sensitive to the detailed structure of the diffractive eigenstates.

\end{abstract}                                                        
\vspace*{0.5cm}                                                    
                                                    
\end{center}

\section {Introduction}
The measurements of diffractive processes obtained at the LHC \cite{TOTEM,ATLAS,CMS,ALICE} are intriguing.  We  summarize two particular unexpected aspects of the data as follows.  First, the $pp$ total  cross section, $\sigma_{\rm tot}$, grows with energy a bit faster than was predicted either from a simple Donnachie-Landshoff parameterization \cite{DL} or from numerous simple theoretical models. This is contrary to the naive expectation that the growth would slow down due to increasing absorptive effects.

On the other hand, the probability of the proton to diffractively dissociate into a relatively low mass state, $N^*$, at the LHC is much less than was expected. Indeed, at fixed-target and CERN-ISR energies cross section for low mass dissociation, \sd, is about 30$\%$ of the elastic cross section \se ~\cite{ABK}. If we were to describe the dissociation just via the (pomeron$-p-N^*$)-vertex\footnote{This vertex factor is denoted $V(p \to N^*)$ below. We use $N^*$ as a generic name for low-mass nucleon resonances and other low-mass excitations.}, then we would expect about the same ratio at 7 TeV; or a little bit less due to stronger absorptive corrections at higher energy. Indeed, in popular models \cite{KMR,TA,Ostap} describing `soft' physics, where low-mass dissociation is included in terms of the Good-Walker (GW) formalism \cite{GW}, the prediction is \sd$\sim7-10$ mb at 7 TeV, whereas TOTEM\footnote{Actually the TOTEM result is based on  the difference between the 
total rate of inelastic events, obtained using optical theorem,
 and the observed rate
 of events with at least one charged particle with $|\eta| < 6.5$.
According to Monte Carlo simulations, this difference corresponds to 
processes where the mass of the dissociating system is less than 3.4 GeV. These processes should mainly originate from the hadronisation of the GW eigenstates. As a rule, particles coming from the fragmentation of a low-mass system $(M_{\rm diss} \sim 2-3$ GeV  produced around mean rapidity $y \sim 8$) are spread out over a $|\eta| \sim 1.5$ rapidity interval, that is, just down to $\eta=6-6.5$ starting from the rapidity $y_p=8.9$ of an incoming 3.5 TeV proton.}
report \sd$=2.62\pm 2.17$ mb (with a 95$\%$ confidence upper limit of 6.31 mb); and $\sigma_{\rm tot} \simeq 98$ mb and \se$\simeq$ 25 mb. 

In this paper, we discuss whether it is possible to describe simultaneously the whole set of high energy diffractive data, including $\sigma_{\rm tot}$, the elastic differential cross section, $d$\se$/dt$, and cross section for low-mass dissociation, \sd, measured at the CERN-ISR \cite{SDisr} and the LHC \cite{TOTEM}, as well as the high-mass dissociation, $d\sigma /d(\Delta\eta)$, measured by ATLAS \cite{ATLAS} and CMS \cite{CMS}. It turns out that this is possible in a framework based on using only one pomeron. However, in order to explain the low value of \sd observed at the LHC we have to take more care about the detailed implementation of the Good-Walker formalism. We describe this next.

\section{Good-Walker formalism}
\subsection{The basic idea}
High energy diffractive interactions are described in terms of the exchange of the rightmost Regge pole, the pomeron, in the complex angular momentum plane. Besides the elastic $p \to p$ vertex, there is the possibility of $p \to N^*$ transitions; that is, the pomeron may distort the wave function of the incoming proton leading to the production of higher nucleon resonances. In the very naive case, with only a single pomeron pole exchange, the probability of $N^*$ excitations is given just by the ratio of the vertices
\be
R=\left( \frac{V(p \to N^*)}{V(p \to p)}\right)^2.
\label{eq:1}
\ee
However, we have to account for multi-pomeron (absorptive)  effects, which within the eikonal model are described by diagrams containing several $t$-channel pomeron exchanges. Clearly, the fact that each vertex may produce one or another $N^*$ resonance complicates the calculation.

In the Good-Walker (GW) paper \cite{GW} it was proposed to first diagonalize the transition matrix in such a way that the interaction of each GW (or so-called diffractive) eigenstate, $\phi_i$, can be described by a simple one-channel eikonal. That is, the eigenstates $\phi_i$ only undergo pure `elastic' scattering. After this the wave function expanded in terms of the $\phi_i$ may be decomposed back into the physical $p$ and $N^*$ states. Since each $\phi_i$ state may have its own interaction amplitude, the outgoing wave function will not coincide with that of the incoming proton. The coherence of the original proton is lost leading to $p \to N^*$ dissociation.

As a rule, multi-channel eikonal models use a set of GW eigenstates such that the eigenstates do not depend on the momentum transfer or the interaction energy. Therefore the probability of the $p \to N^*$ excitation does not differ too much from the estimate given in (\ref{eq:1}). Only at very high energy, when the black disc limit is approached, will absorptive corrections strongly suppress dissociation, since a black disc completely absorbs all $\phi_i$ eigenstates. Since, at LHC energies, we approach the black disc limit only at the centre (that is, impact parameter $b$=0), the predicted value of \sd at 7 TeV is not much smaller than the naive estimate.

\subsection{A more physical GW decomposition \label{sec:2.2}}
The GW decomposition may not be so trivial. First, clearly the transition vertex will depend on the momentum transfer squared $t$. Recall that at fixed-target and ISR energies the $pp$ elastic slope $B\sim10~{\rm GeV}^{-2}$ \cite{abk,B10}, whereas at the LHC it is observed to be $B\sim20~{\rm GeV}^{-2}$ \cite{TOTEM}.  That is the structure of the `mean transition matrix' may vary with energy. A larger value of $|t|$ at lower energies will correspond to a more strongly distorted proton wave function, and will lead to a larger probability of $N^*$ excitation. Besides this, at the lower (fixed target) energies there may be excitations due to secondary Reggeon exchange.

Another complication is that at high energy we never deal with pure pomeron pole exchange, but instead mainly with pomeron `cuts'. That is the properties of the `effective' pomeron change with energy. First, recall that already in leading log BFKL \cite{book}, the vacuum singularity (the pomeron) is not a pole, but a cut. Due to the diffusion in log($k_T$) space, the typical transverse momentum inside the pomeron slowly increases with energy. Moreover the (semi-enhanced) absorptive corrections suppress the low $k_T$ contribution ($\sigma_{\rm abs}\propto 1/k^2_T$), and since the absorptive effects become stronger at high energy \cite{KMRJPhysG}, the mean $k_T$ increases. How will this effect the $\phi_i$ cross sections?

At high energies a good example of a GW eigenstate is a state formed by valence quarks, whose position in the impact parameter ($b$) plane is fixed; the interaction with the QCD pomeron (that is, with two $t$-channel gluons) does not change the $b$ coordinates. Thus, let us start with the simple two-gluon Low-Nussinov \cite{LN} pomeron exchange.  In this case, the cross section for pomeron exchange between two dipoles is given by
\be
\sigma_{ab}~=~\int \frac{dk_T^2}{k_T^4}~ \alpha^2_s~[1-F_a(4k_T^2)]~[1-F_b(4k_T^2)].
\ee 
Here the infrared divergency at small $k_T$ is cutoff by the interaction with the quark spectators. In the simplified dipole model this effect is described by the factors [...] in the numerator, where $F_i(4k^2_T)$ are the form factors of the incoming colourless dipoles. Due to this cutoff, the cross section $\sigma_{ab} \propto \alpha^2_sr^2$. That is, a larger size GW  component corresponding to a  larger $r$, has a larger cross section.

If, on the contrary, the integral is cutoff at a larger $k_T$ by some $k_{\rm min}$ arising from the internal structure of the effective pomeron (in a region where the $F_i(4k^2_T)\ll 1)$, then the cross sections of the different GW components will be practically the same.
That is, the value of the cross section is specified by the cutoff induced by the pomeron, and not by the size of the GW eigenstates. As a consequence, all eigenstates have the same cross section, so there is no dispersion, and the interaction will not destroy the coherence of  the wave functions of the incoming protons. Hence, the probability of diffractive dissociation will be negligible. As was discussed in \cite{KMRJPhysG}, the value of $k_{\rm min}$ increases with energy. This behaviour was shown theoretically in \cite{KMR}, and phenomenologically it was observed in the tuning of the Pythia8 Monte Carlo \cite{P8}, where the cutoff has the behaviour
\be
k^2_{\rm min}~\sim ~s^{\beta} ~~~~~~ {\rm where} ~~~~~~~~\beta\simeq 0.24.
\ee

The above two different constructions of the GW eigenstates therefore have quite distinct expectations for the cross section for low-mass diffractive dissociation, \sd,  at the LHC. The simple, conventional approach leads to a marked growth of \sd with energy, due to the growth of the pomeron exchange amplitude as $s^{(\alpha_P(0)-1)}$. However, the approach based on the observed growth of $k_{\rm min}$ with energy gives a much lower value of \sd at the LHC, since simultaneously the dispersion between the cross sections of the $\phi_i$ eigenstates decreases with energy. We illustrate these behaviours by fitting to all the diffractive data using four different implementations of the GW eigenstates. Two for each of the above two formalisms,

\subsection{Description of diffractive data by the GW formalism  \label{sec:2.3}}

To explore the sensitivity to the different constructions of the GW eigenstates, we tune the different approaches to best describe the diffractive data in the CERN-ISR to LHC energy range. To be precise we include the measurements of $\sigma_{\rm tot},~d$\se$/dt, $ \sd and $d\sigma/d(\Delta \eta) $. The expressions for the observables are given in terms of the GW eigenstates in the Appendix. In each case we use a two-channel eikonal, $i,k=1,2$, and parametrise the form factor of each state in the form
\be
F_i(t)={\rm exp}(-(b_i(c_i-t))^{d_i}+(b_ic_i)^{d_i}),
\label{eq:ff}
\ee
where $c_i$ is added to avoid the singularity $t^{d_i}$ in the physical region of $t<4m^2_{\pi}$. Note that $F_i(0)=1$. The six parameters $b_i,~c_i,~d_i$, together with the intercept and slope of the pomeron trajectory are tuned to describe the elastic scattering data, paying particular attention to the energy behaviour of \sd.

First, we study two fits based on the simple, conventional implementation of the GW formalism. In fit 1 we tune the parameters  of the GW eigenstates to reproduce \sd =2 mb at the CERN-ISR energy (close to the lower experimental bound) and to give the smallest possible value of \sd ($\sim 5$ mb) at 7 TeV. In fit 2, we require \sd =1 mb at the CERN-ISR energy, assuming that another 1 mb arises from a secondary Reggeon contribution and from a larger distortion of the incoming proton wave function due to the larger momentum transfer (recall the lower elastic slope $B$ at the lower energy). In such a fit we find \sd$\simeq2.8$ mb at 7 TeV, compatible with the TOTEM observations. In both of these fits there is no energy dependence of the pomeron $k_T$.

The possible role of an energy dependent $k_{\rm min}$ is studied in fits 3 and 4. To mimic the effect discussed at the end of Subsection \ref{sec:2.2}, we write the cross section for the interaction of GW eigenstate $\phi_i$ and $\phi_k$, via one-pomeron-exchange in the form
\be
\sigma_{ik}=\sigma_0\gamma_i\gamma_k(s/s_0)^\Delta,
\ee
where $\Delta$ is the `intercept' of the pomeron. More precisely the pomeron has trajectory\footnote{Besides the constant slope, we insert the $\pi$-loop contribution as proposed in \cite{AG}, implemented as in \cite{KMR18}}
\be
\alpha_P(t)=1+\Delta+\alpha'_P t.
\ee
Since $\sigma (\phi_i) \equiv \sigma_{ii}\propto \alpha_s^2r_i^2 \propto \gamma_i^2$, we parametrize $\gamma_i$ in the form
\be
\gamma_i~ \propto~ \frac{1}{1+k_i/k_{\rm min}(s)},
\label{eq:gamma}
\ee
which, at low energies where $k_{\rm min}$ is small, gives some non-trivial value $\gamma_i \propto k_{\rm min}/k_i$; but which, for large $k_{\rm min}$, tends to $\gamma_i =1$.

In the extreme case, fit 3, we take $k_{\rm min}^2\propto s^{0.24},$ corresponding to the behaviour found in tuning the Pythia8 Monte Carlo. However, this value of $k_{\rm min}$ is appropriate for the central rapidity region, while dissociation occurs in the proton fragmentation domains. Therefore, in fit 4, we consider a less steep energy behaviour, 
\be
k_{\rm min}^2 \propto s^\beta ~~~~~~{\rm with}~~~~~~\beta=0.12.
\ee

 \section{High-mass dissociation}
\begin{figure} [htb]
\begin{center}
\vspace*{-5.0cm}
\includegraphics[height=15cm]{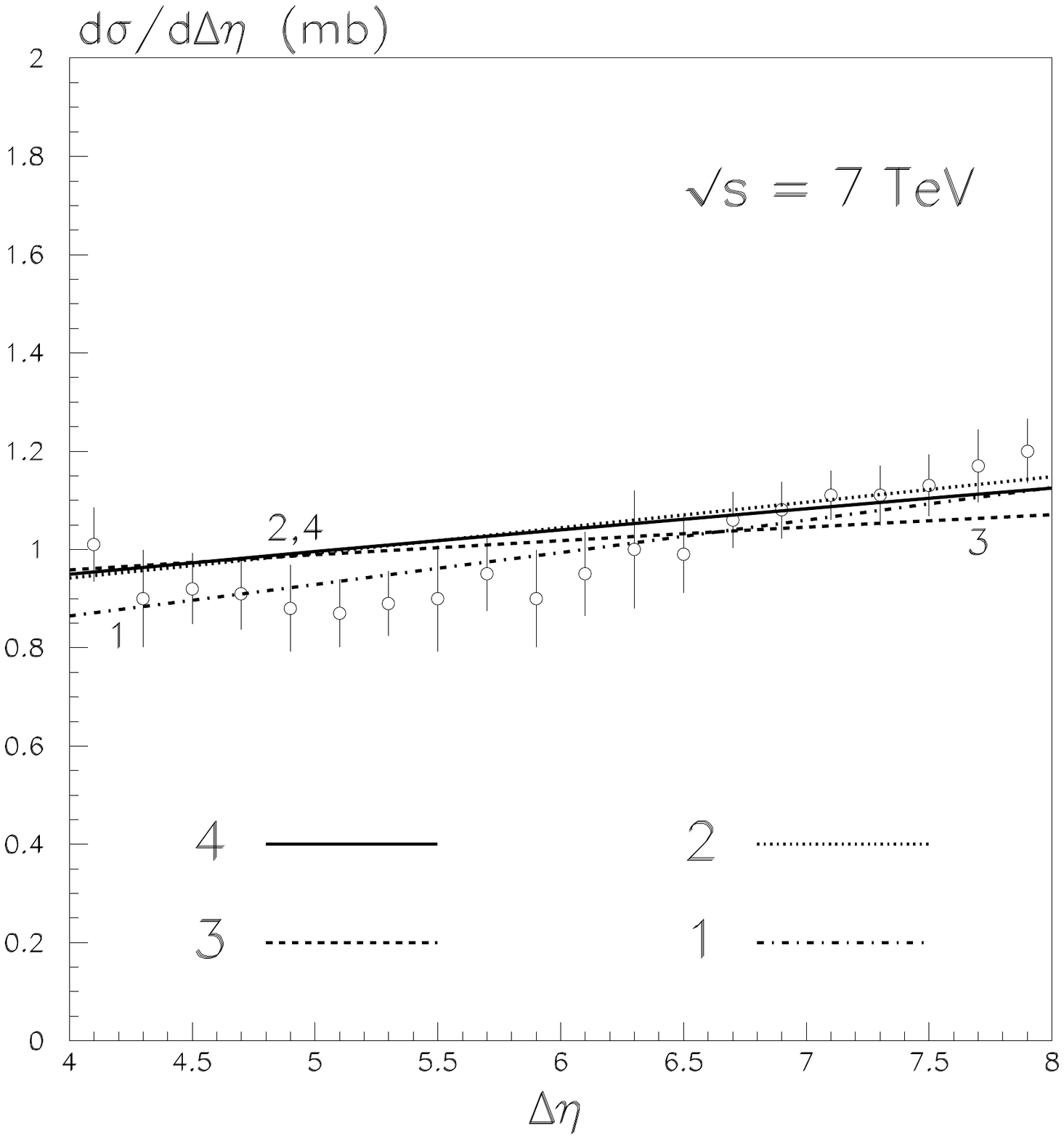}
\vspace{-0.0cm}
\caption{\sf The ATLAS \cite{ATLAS} measurements of the inelastic cross section differential in rapidity gap size $\Delta\eta$ for particles with $p_T>200$ MeV. Events with small gap size ($\Delta\eta \lapproxeq 5$) may have a non-diffractive component which arises from fluctuations in the hadronization process \cite{KKMRZ}. This component increases as $\Delta\eta$ decreases (or if a larger $p_T$ cut is used \cite{KKMRZ,ATLAS}).  
The data with $\Delta\eta \gapproxeq 5$ are dominantly of diffractive origin, and may be compared with predictions of the 4 models. }
\label{fig:eta}
\end{center}
\end{figure}

 So far high-mass, $M$, dissociation at the LHC has not actually been measured as a $M^2d\sigma/dM^2$ distribution, which is usually used in Regge theory as described in Appendix B. Instead both the ATLAS \cite{ATLAS} and CMS \cite{CMS} collaborations have selected large rapidity gap events using information from the inner detector tracks and the calorimeter in a large rapidity interval around $|\eta|=0$. For example, the ATLAS experiment \cite{ATLAS}
 detects particles in the rapidity interval $|\eta|<4.9$, while the larger rapidity interval up to proton $y=\pm 8.9$ is uninstrumented.  ATLAS measure $d\sigma/d(\Delta\eta)$ with $\Delta\eta$ defined by the larger of the two empty $\eta$ regions extending between the edges of the detector acceptance at $\eta=4.9$ or $\eta=-4.9$ and the nearest track or cluster, passing the selection requirements, at smaller $|\eta|$. The gap size relative 
to $\eta=\pm 4.9$ lies in the range $0<\Delta\eta<8$, such that, for example, $\Delta\eta=8$ implies that
no final state particles are produced above a transverse momentum threshold
$p_T^{\rm cut}=200$ MeV in one of the regions $-4.9<\eta<3.1$ or $-3.1<\eta<4.9$.   We compare our predictions for $d\sigma/d(\Delta\eta)$ with the data, using an analogous procedure to that developed in \cite{KMRopacity}. The results are shown in Fig. \ref{fig:eta} for the four versions of the GW eigenstates found in Section \ref{sec:2.3}.

\section{Gap survival factors}
\begin{figure} 
\begin{center}
\vspace{-4cm}
\includegraphics[height=12cm]{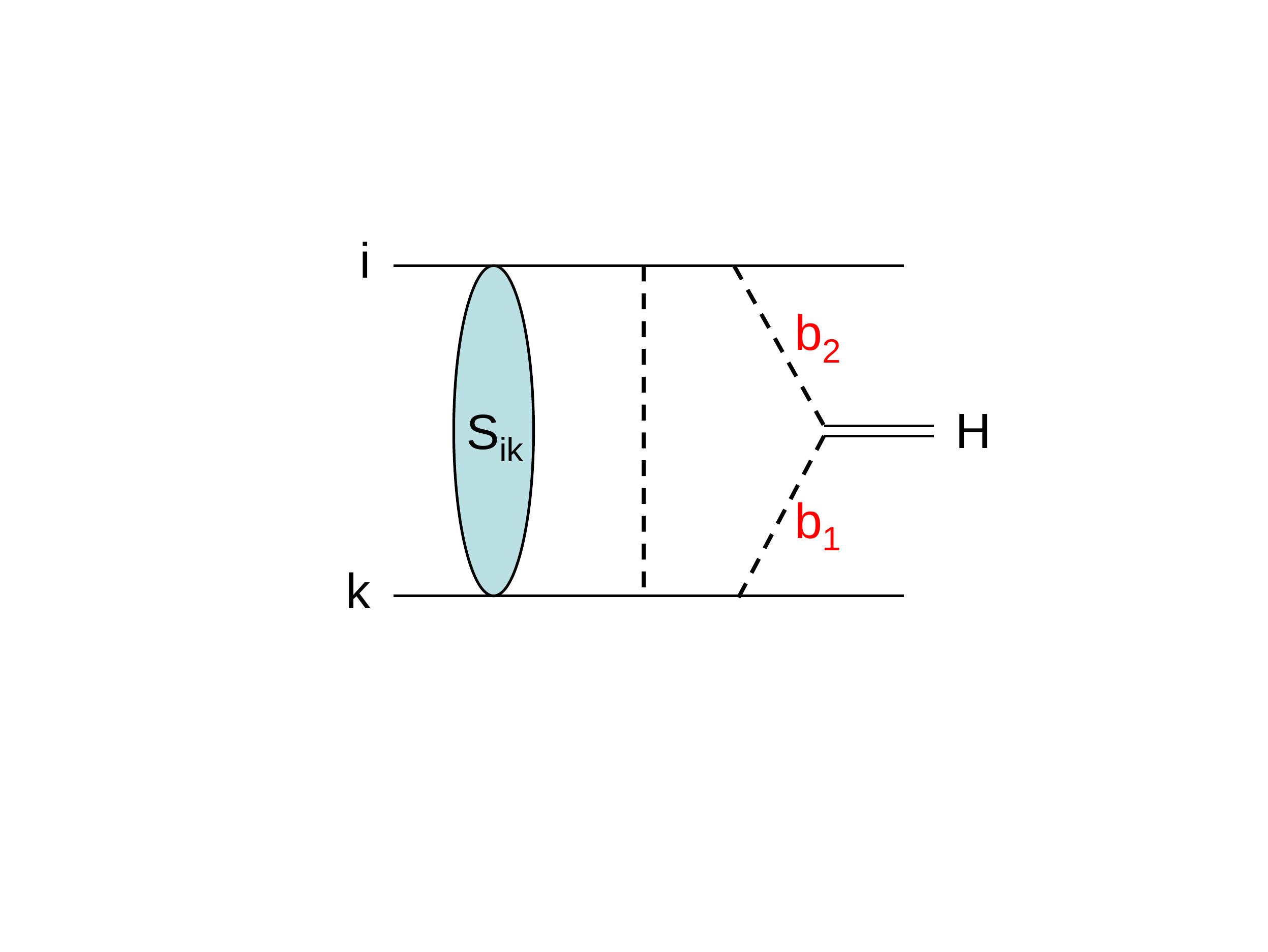}
\vspace*{-4.5cm}
\caption{\sf The diagram describing the amplitude of exclusive Higgs boson production, $ik \to p+H+p$, where $i,k$ are GW diffractive eigenstates. The dashed lines are gluons. We also include eikonal absorptive corrections which lead to a survival factor $S_{ik}~=~{\rm exp}(-\Omega_{ik}/2)$ of the rapidity gaps either of the Higgs boson.}
\label{fig:higgs}
\end{center}
\end{figure}

To calculate the cross sections of low multiplicity exclusive processes at high energies, it is important to know the gap survival factors. That is, the probability that extra secondaries which may be produced in additional (multiple) interactions between the spectators do not populate the rapidity gaps. In other words, do not spoil the exclusivity of the process.  The major suppression comes from the interaction of the incoming parton spectators, which within the eikonal model, is described by the multiple rescattering, shown symbolically as $S_{ik}$ in Fig. \ref{fig:higgs}.  For illustration, we consider the survival factor $S^2$ for exclusive Higgs production, $pp \to p+H+p$, where the + signs denote large rapidity gaps.

To calculate the cross section for $pp \to p+H+p$ we work in impact parameter space\footnote{Here we neglect the size, $\Delta b_h$, of the `hard matrix element' in comparison with the size of the proton. That is, we assume that the amplitude of the hard subprocess is point-like in $b$-space.}. The cross section, as a function of the rapidity of the Higgs boson, has a somewhat similar structure to (\ref{eq:result}) in Appendix B
\be
\frac{d\sigma}{dy}~=~N\left|\sum_{i,k}|a_i|^2|a_k|^2\int\frac{d^2b_2}{\pi}\int\frac{d^2b_1}{2\pi} \Omega_i(b_2,y) \Omega_k(b_1,Y-y) \cdot S_{ik}(\vec{b}_2-\vec{b}_1)\right|^2,
\label{eq:php}
\ee
where $N$ is normalisation factor, $Y={\rm ln}(s/m_p^2)$, 
\be
S_{ik}~=~{\rm exp}(-\Omega_{ik}/2),
\label{eq:Sik}
\ee
 and $\Omega_i(\Omega_k)$ corresponds to the opacity of the state $i(k)$ probed by the corresponding active incoming parton in the hard subprocess, We assume the opacities are described by the same effective pomeron. However, since here we consider partons at large scale we put the slope of the pomeron trajectory $\alpha'_P=0$.
 
 In comparison with previous estimates of the survival factor \cite{KMRsurvival}, we now anticipate a stronger suppression, that is a smaller value of
  \be
 \langle S^2 \rangle~=~\frac{\int d^2b_2\int d^2 b_1
\left|\sum_{i,k}|a_i|^2|a_k|^2 ~\Omega_i(b_2,y) \Omega_k(b_1,Y-y) \cdot
S_{ik}(\vec{b}_2-\vec{b}_1)\right|^2}{\int d^2b_2\int d^2 b_1
\left|\sum_{i,k}|a_i|^2|a_k|^2~ \Omega_i(b_2,y) \Omega_k(b_1,Y-y)\right|^2},
 \ee
 since previous models of soft phenomena underestimate the total cross section at the LHC. The dependence on $\sigma_{\rm tot}$ is very strong, as it enters as an exponent, see (\ref{eq:Sik}) and (\ref{eq:tot}). On the other hand, the value of $\langle S^2 \rangle$ is very sensitive to the detailed structure of the GW eigenstates; that is to the probability to find an active parton at a particular $b$ value in one or another eigenstate, and to the $b$ distributions of these partons.

We give the values of the survival factor $S^2$ for the exclusive production of a heavy object, integrated over the transverse momenta of the recoil protons, (for example, for exclusive Higgs boson production, $pp \to p+H+p$) for all 4 versions of the model in the last column Table \ref{tab:A2}.  The values of $S^2$ vary noticeably from one version to another, despite the fact that all versions are in good agreement with the elastic LHC data. In particular, at 7 TeV, the value of $S^2$ in version 2, which has the smaller \sd, is {\it three} times greater than that in version 1.  A larger \sd means a stronger dispersion (a larger difference) between the GW eigenstates, and hence a stronger screening. Indeed, given that the gluon PDF is proportional to the coupling of the corresponding eigenstate, we have the major contribution from the state with the largest cross section.  On the other hand, the probability of an additional inelastic interaction in this state is also larger, leading to a smaller survival factor $S^2$.

Another interesting observation is that the value of $S^2$ is  sensitive to the behaviour of the form factors of the GW eigenstates, $F_i(t)$ of (\ref{eq:ff}); in particular to the spatial distribution in impact parameter, $b$, space.  The gap survival probability practically nullifies the possibility of getting exclusive production from the centre of the disc. The main contribution to the process comes from the periphery, and thus depends strongly on the shape of the tail in $b$-space, that is on the behaviour of the form factors, $F_i(t)$ of (\ref{eq:ff}). In particular, two equally good descriptions of the elastic $pp$ scattering data, with the same values of \sd, may easily give 30$\%$ difference in the value of $S^2$.

 \section{Details of the models and description of data}
 Recall that we are using a two-channel eikonal, and so we have two GW diffractive eigenstates.  The parameters of the model are listed in the first column of Table \ref{tab:1}. We have already introduced many of them in Section \ref{sec:2.3}. Here, we give a more detailed discussion of the parameters, particularly discussing the energy dependence arising from that of $k_{\rm min}(s)$, which distinguishes versions 3 and 4 of the model from versions 1 and 2. 
  
  The first seven rows of the Table show the values of the parameters of the pomeron trajectory, $\alpha_P(t)$, and the pomeron couplings. The coupling, $v_i$, to each GW eigenstate is presented in terms of the average cross section of the eigenstates: $\sigma_0\equiv (\sigma (\phi_1)+\sigma (\phi_2)/2$; that is
 \be
 \sigma~=~\sigma_0(s/1~{\rm GeV}^2)^{\alpha_P(0)}, ~~~~~~~{\rm with}~~~~~~~ {v}_i=\sqrt{\sigma_0}~\gamma_i.
 \ee
 So the parameters $\gamma_i$ are dimensionless, and 
 \be
 (\gamma_1+\gamma_2)/2=1.
 \ee
 
 In versions 1 and 2 we take
 \be \
 \gamma_{1,2}~=~1\pm \gamma,
 \label{eq:nor}
 \ee
 where $\gamma$ is a parameter.  On the other hand in versions 3 and 4 the values of the $\gamma_i$ depend on the energy. Following (\ref{eq:gamma}), we take
 \be
\gamma_{1,2}~=~1\pm\frac{k_2-k_1}{k_{\rm min}(s)+(k_1+k_2)/2}.
\label{eq:gamma2}
\ee
In comparison with (\ref{eq:gamma}), here we account for the normalisation given by (\ref{eq:nor}). So in these versions of the model, the parameters are now the $k_i$, which characterize the momenta of the eigenstates $\phi_i$.
  
 \begin{table}[htb]
\begin{center}
\begin{tabular}{|l|c|c|c|c|}\hline
 &   1 &  2 &   3 & 4  \\ \hline
  $\Delta$ & 0.13 & 0.115 & 0.093 & 0.11 \\
  $\alpha'_P~(\GeV^{-2})$ & 0.08 & 0.11 & 0.075 & 0.06 \\
  $\sigma_0 $ (mb) & 23 & 33 & 60 & 50\\
  $\lambda$(1.8 TeV) & 0.2 & 0.17 & 0.19 & 0.19 \\
  $\gamma$ & 0.55 & 0.4 & - & - \\
  $k_1/k(1.8$ TeV) & - & - & 1.03 & 1.3 \\
    $k_2/k(1.8$ TeV) & - & - & 4.8 & 6.0 \\
  \hline
  $|a_1|^2 $ & 0.46 & 0.25 & 0.24 & 0.25 \\
  $ b_1~(\GeV^{-2})$  & 8.5 & 8.0 & 5.3 & 7.2 \\
  $ c_1~(\GeV^2) $ & 0.18 & 0.18 & 0.35 & 0.53 \\
  $d_1$ & 0.45 & 0.63 & 0.55 & 0.6\\
  $ b_2~(\GeV^{-2})$  & 4.5 & 6.0 & 3.8 & 4.2 \\
  $ c_2~(\GeV^2) $ & 0.58 & 0.58 & 0.18 & 0.24 \\
  $d_2$ & 0.45 & 0.47 & 0.48 & 0.48\\
 \hline

\end{tabular}
\end{center}
\caption{\sf The values of the parameters in the four versions of the two-channel eikonal fit to elastic $pp$ scattering data in which particular attention is paid to the value of \sd and to the behaviour of the GW eigenstates. The first seven rows give the values of parameters connected to the pomeron trajectory and its couplings, and the last seven rows list the parameters which specify the GW eigenstates.}
\label{tab:1}
\end{table}

 The triple-pomeron coupling is written in the form
 \be
 g_{3P}~=~\lambda~g_N
 \ee
 where, $g_N$ is the pomeron-proton coupling. In versions 1 and 2 of the model, $\lambda$ is a simple parameter independent of the energy, whereas it is taken to have the energy-dependent form in versions 3 and 4, similar to that in  (\ref{eq:gamma}) and (\ref{eq:gamma2}), 
 \be
 \lambda=\frac{1}{1+k_{\rm min}(s)/k_{3P}},
 \ee
 where $k_{3P}$ is the parameter in these latter two models.  It turns out that at the Tevatron energy the corresponding value of $\lambda$ is about 0.19 for versions 3 and 4; that is, essentially equal to the energy independent value found in versions 1 and 2, and, moreover, in agreement with the value found in the triple-Regge analysis \cite{luna} of Tevatron and lower energy data.
 
 The last seven rows of Table \ref{tab:1} list the values of the parameters which describe the detailed structure of the two GW eigenstates. The first entry gives the value of $|a_1 |^2$, while $|a_2 |^2=1-|a_1 |^2$. Recall that $|a_i|^2$  is the probability to find eigenstate $\phi_i$ in the proton. see (\ref{eq:ai}). The other parameters specify the form factors of the eigenstates, see (\ref{eq:ff}).
 
 \begin{figure} 
\begin{center}
\vspace*{-1.cm}
\includegraphics[trim = 0mm 0mm 0mm 100mm, clip, height=11cm]{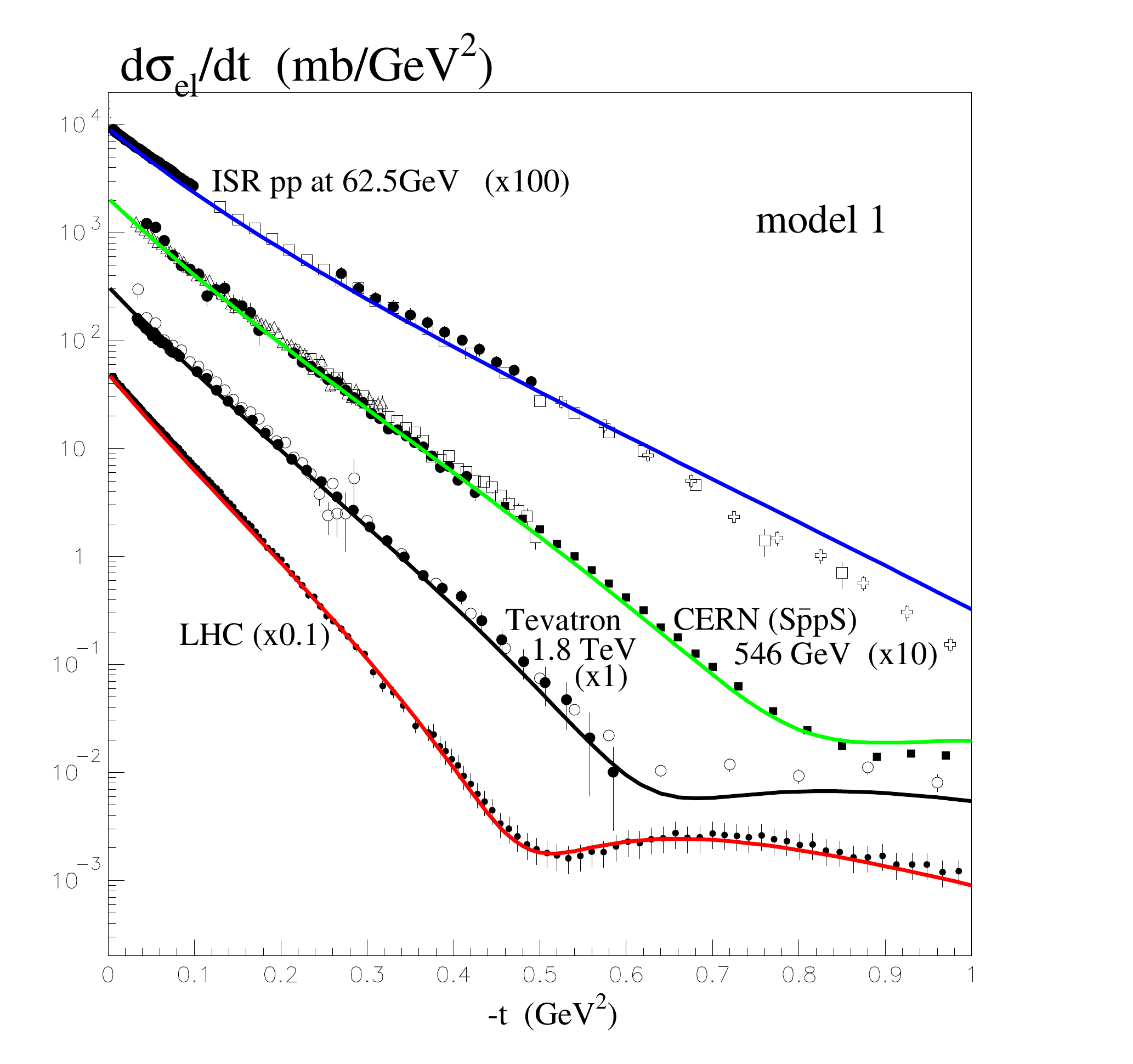}
\includegraphics[trim = 0mm 0mm 0mm 100mm, clip,height=11cm]{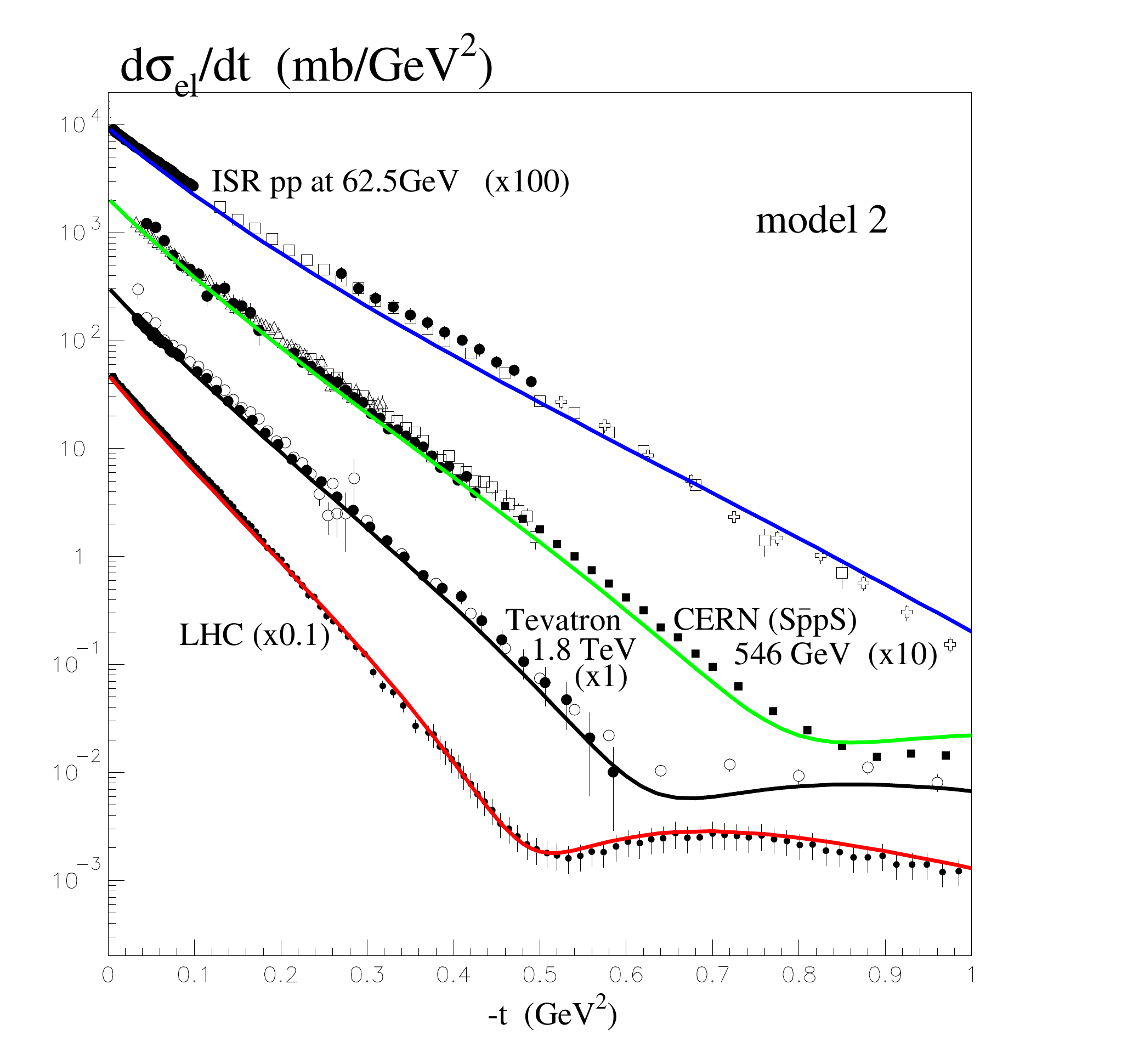}
\vspace*{-0.5cm}
\caption{\sf The description of $pp$ or ($p{\bar p}$) elastic data in models 1 and 2, respectively. If $\sqrt{s}=62.5 ~\GeV \to~$7 TeV, then \sd $\simeq2 \to 5$ mb in model 1, and \sd $\simeq 1 \to 2.8$ mb in model 2. The data are taken from \cite{elastic}. Here LHC refers to 7 TeV.}
\label{fig:s1}
\end{center}
\end{figure}
\begin{figure} 
\begin{center}
\includegraphics[trim = 0mm 0mm 0mm 100mm, clip,height=11cm]{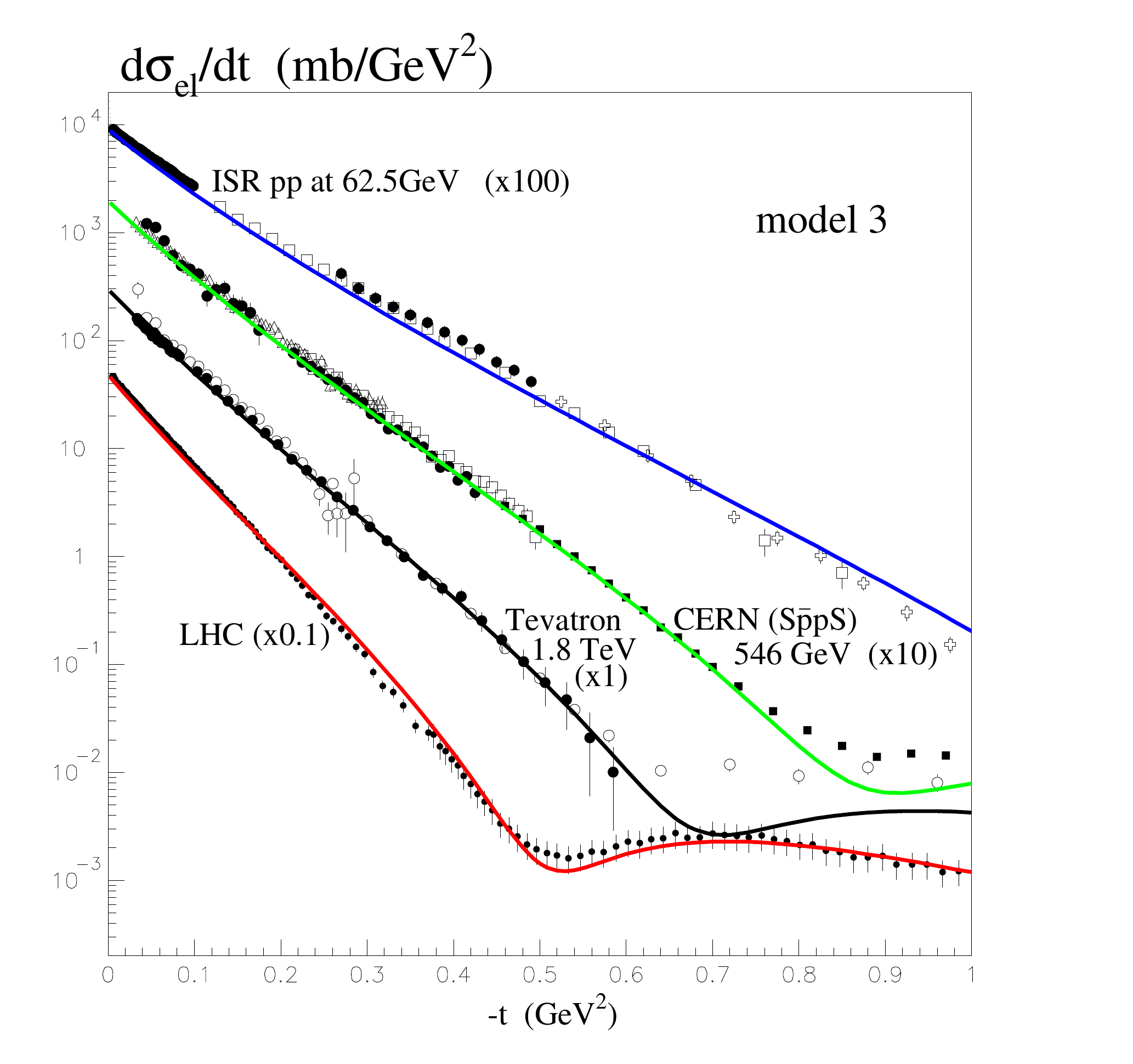}
\includegraphics[trim = 0mm 0mm 0mm 100mm, clip,height=11cm]{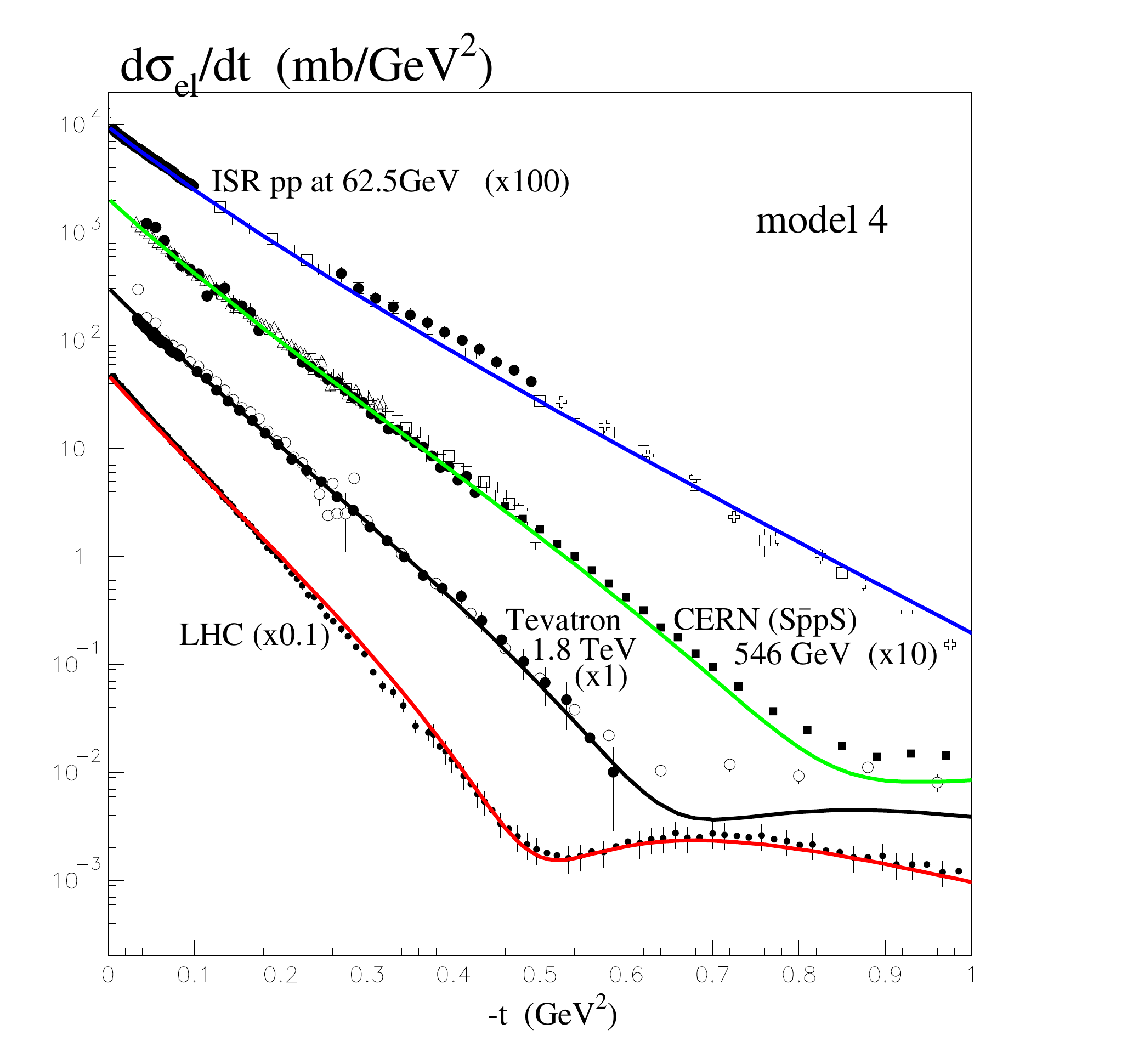}
\vspace*{-0.5cm}
\caption{\sf The description of $pp$ or ($p{\bar p}$) elastic data in models 3 and 4, respectively, which correspond to $k_{\rm min}^2(s)\propto s^{0.24}$ and $k_{\rm min}^2(s) \propto s^{0.12}.$  The data are taken from \cite{elastic}. Here LHC refers to 7 TeV.}
\label{fig:s3}
\end{center}
\end{figure}

Finally, to calculate the elastic cross section, $d\sigma_{\rm el}/dt$ we include the real part of the amplitude.  This contribution is crucial in the region of the diffractive dip. The real part is computed using a dispersion relation. For an even-signature amplitude
\be
A~\propto ~s^{\alpha} + (-s)^{\alpha} ~~~~~~~{\rm we ~have} ~~~~~~~\frac{{\rm Re}~A}{{\rm Im}~A}={\rm tan}(\pi\alpha /2),
\label{eq:RE}
\ee 
 that is the usual signature factor. This formula is transformed into $b$-space, so that the complex opacities, $\Omega_{ik}(b)$ in (\ref{eq:dsel}) can be constructed. For each value of $b$, that is for each partial wave $\ell$, we calculate $\alpha$ and determine Re$~A$ from (\ref{eq:RE}).

 \begin{table}[!phtb]
\begin{center}
\begin{tabular}{|c|c|c|c|c|c|c|c|c|}\hline
 &  $\sqrt{s}$ & $\sigma_{\rm tot}$  & $\sigma_{\rm el}$  & $B$  &   $\sigma^{\rm SD}_{{\rm low}M}$   &  $\sigma^{\rm DD}_{{\rm low}M}$   &     \sd     &     $S^2$ \\
 &  TeV &  mb &  mb & $\GeV^{-2}$ & mb & mb & mb &   \\
 \hline
  &  0.0625  & 42.0   &  6.8  &   13.3  &   {\bf 2.02}  &  0.14 &    2.16   &    0.105\\
  &   0.546  & 63.1   &  12.5  &   16.2  &   3.14  &  0.22&    3.36   &    0.041\\
model 1 &   1.8  & 77.6   &  16.9  &   18.2  &   3.85  &  0.28 &    4.13   &    0.023\\
& 7 &    97.0 &     23.2 &    20.7  &   4.72 &   0.37  &   {\bf 5.09}   &    0.011  \\
 &  14  & 108   &  26.9  &   22.1  &   5.20  &  0.42 &    5.61   &    0.007\\
 & 100  & 144   &  39.6  &   26.7  &   6.64  &  0.57  &   7.21   &    0.002\\
\hline
 &  0.0625  & 42.2  &   6.6  &   13.9  &  {\bf 1.00}  &  0.03& 1.03    &   0.208\\
  &   0.546  & 62.8  &   12.1  &   16.6  &   1.67  &  0.05& 1.72   &    0.103\\
model 2 &   1.8  & 77.1  &   16.5  &   18.4  &   2.13  &  0.07& 2.20  &     0.063\\
 &   7  & 96.1  &   22.8  &   20.7  &   2.71  &  0.09& {\bf 2.81}   &    0.032\\
 &  14  & 107  &   26.6  &   22.0  &   3.04  &  0.11    & 3.14   &    0.022\\
&  100  & 143  &   39.5  &   26.2  &   4.02  &  0.16    & 4.18   &    0.006\\
 \hline
  &  0.0625  & 41.7   &  6.7  &   13.4  &   {\bf 1.99}  &  0.10 &    2.09   &    0.087\\
  &   0.546  & 61.5   &  12.0  &   16.1  &   2.28  &  0.10&    2.38   &    0.047\\
model 3 &   1.8  & 76.1   &  16.5  &   17.9  &   2.32  &  0.09 &    2.41   &    0.031\\
& 7 &    96.6 &     23.5 &    20.3  &   2.24 &   0.07  &   {\bf 2.31}   &    0.017  \\
 &  14  & 109   &  27.8  &   21.7  &   2.14  &  0.06 &    2.21   &    0.012\\
 & 100  & 149   &  43.3  &   26.3  &   1.74  &  0.03  &   1.77   &    0.004\\
\hline
 &  0.0625  & 42.8  &   7.1  &   13.1  &  {\bf 2.02}  &  0.11 & 2.13    &   0.100\\
  &   0.546  & 63.0  &   12.9  &   15.7  &   2.57  &  0.13 & 2.69  &    0.047\\
model 4 &   1.8  & 77.2  &   17.4  &   17.5  &   2.82  &  0.14& 2.95  &     0.028\\
 &   7  & 96.4  &   24.0  &   19.8  &   3.05  &  0.14& {\bf 3.19}  &      0.015\\
 &  14  & 108  &   27.9  &   21.1  &   3.15  &  0.14   & 3.29   &    0.010\\
&  100  & 145  &   41.8  &   25.5  &   3.33  &  0.14    & 3.47   &    0.003\\
 \hline
\end{tabular}
\end{center}
\caption{\sf The results obtained from tuning the parameters, of the 4 versions of the 2-channel eikonal model, to describe the elastic $pp$ and $p\bar{p}$ data.  $B=\sigma_{\rm tot}^2/16\pi\sigma_{\rm el}$ is the mean elastic slope, that is  $d\sigma_{\rm el}/dt \sim e^{Bt}$.  The dissociation cross sections shown in bold-face type are those for which experimental measurements exist: at the CERN-ISR we have $\sigma^{\rm SD}_{{\rm low}M} \simeq$ 2$-$3 mb \cite{SDisr}, while at the LHC, TOTEM report \cite{TOTEM} a value \sd=2.6$\pm$2.2 mb at 7 TeV, where \sd is the sum of single dissociation of both protons and double proton dissociation..}
\label{tab:A2}
\end{table}
For each of the 4 versions of the two-channel eikonal model (with absorptive corrections), we tune the parameters to describe the $pp$ (and $p\bar{p}$) elastic scattering data \cite{elastic}.  The values of the parameters are listed in Table \ref{tab:1}, and the description of the elastic data are shown in Figs. \ref{fig:s1} and \ref{fig:s3}.

For completeness, we show in Fig. \ref{fig:larget} the description of the 7 TeV elastic data {\cite{TOTEM} out to larger $|t|$ values, together with the predictions at 14 and 100 TeV, using model 4.
\begin{figure} 
\begin{center}
\vspace*{-2.cm}
\includegraphics[trim = 0mm 0mm 0mm 100mm, clip,height=11cm]{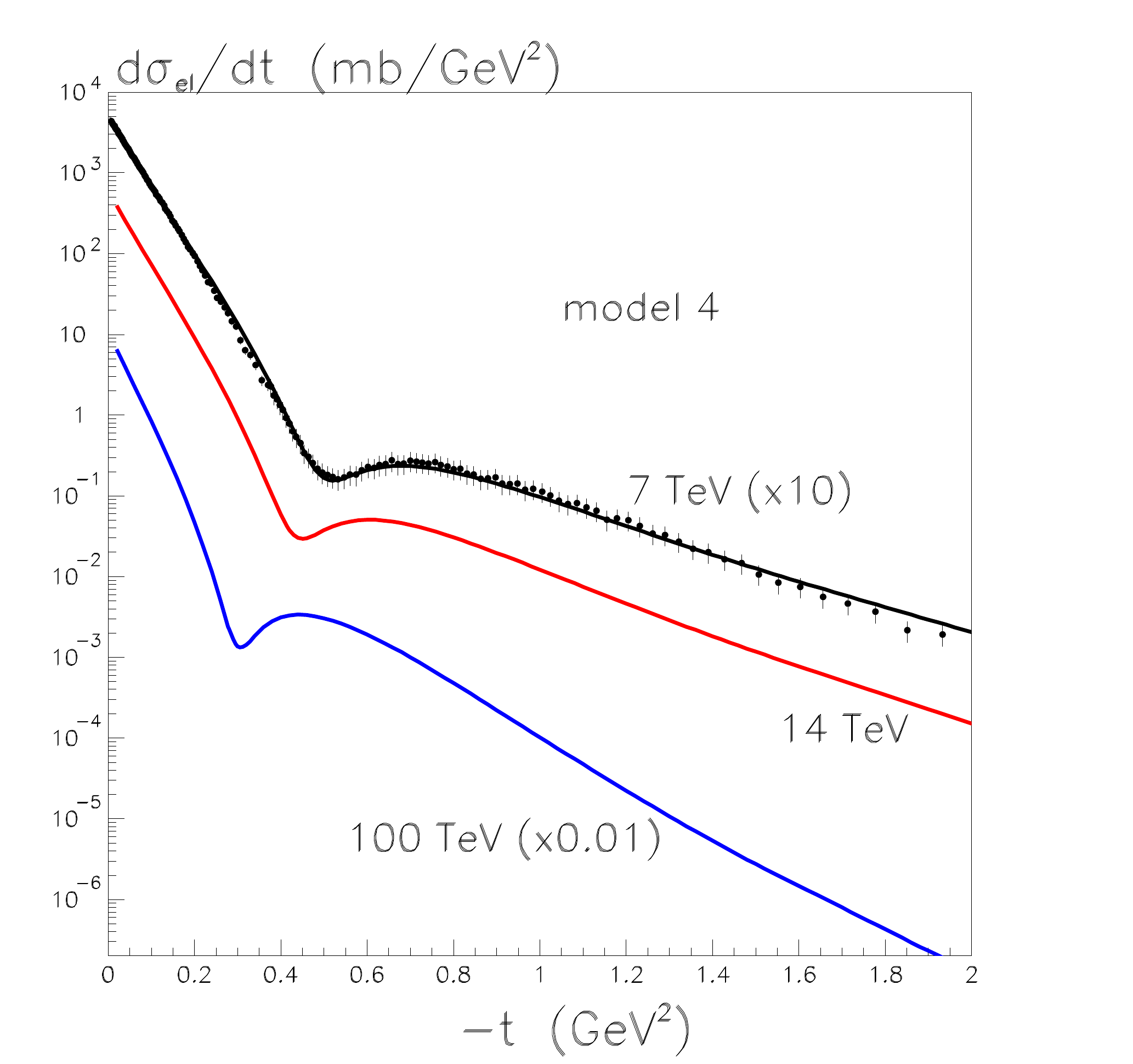}
\caption{\sf The description of the TOTEM $d\sigma_{\rm el}/dt$ data at 7 TeV in model 4, together with the predictions at 14 and 100 TeV.}
\label{fig:larget}
\end{center}
\end{figure}

\section{Discussion and Conclusions}
In all four versions of the model, we see that it is possible to satisfactorily describe $d\sigma_{\rm el}/dt$ for $-t<1 ~\GeV^2$ in the energy range from CERN-ISR to the LHC, and to account reasonably well for the diffractive dip structure, see Figs. \ref{fig:s1} and \ref{fig:s3}. The tuning of the model to describe the data may be a little improved if we were to allow a secondary Reggeon contribution, which is not completely negligible at the lowest (CERN-ISR) energy. However this would almost double the number of parameters, and would not anyway change our conclusions.

It was important to include the real part of the elastic amplitude to describe the data in the region of the diffractive dip. It is amusing to note that to reproduce the dip we found it necessary to parametrise the form factors of the GW eigenstates with a form close to that proposed many years ago by Orear \cite{Orear}, $F(t) \propto {\rm exp}(-\sqrt{|t|})$, see the values of the parameters $d_i$ in Table \ref{tab:1}. On the other hand, there is no theoretical evidence that the form factor should have a Gaussian form. 

 In order to describe, not only the elastic cross section, but also low-mass diffractive dissociation, we use the GW formalism in a two-channel eikonal model. A one-channel eikonal is clearly not adequate, since it gives zero diffractive dissociation.
 
 At first sight, it was unexpected that low-mass dissociation measured at the LHC (\sd$=2.6\pm 2.2$ mb) was found to be practically the same as that measured at much lower CERN-ISR energy (\sd$\simeq 2-3)$ mb, while the elastic cross section increase by more than a factor of 3 in this energy interval. The values found in the various models for these two measurements are highlighted bold-face type in Table \ref{tab:A2}. We have shown that this phenomena may be described 
 \begin{itemize}
 \item {\bf either} by assuming that about half of \sd at the ISR was due to secondary Reggeon contributions and/or  contributions from relatively large $|t|$, which die out at LHC energies (model 2),
 \item {\bf or} by allowing for the coupling of the pomeron to the GW eigenstates, $\gamma_i$ to depend on collider energy, as expected theoretically (models 3,4).
 \end{itemize}
 
 Let us recall why the energy dependence is theoretically expected.
 Note that the vacuum singularity (in the complex angular momentum plane) in QCD is not a pure pole, but a cut. There is BFKL diffusion in log$k_T$ space, which leads to a growth of the typical $k_T$ inside the `pomeron' with energy. In this way it is possible to explain the energy behaviour of \sd. It is not surprising that the value $\Delta \simeq 0.11$ found for the effective pomeron is {\it larger} than the 0.08 (the value obtained when the amplitude was parametrized by one-pole-exchange without any multi-Pomeron corrections
 \cite{DL}), but is {\it smaller} than the intercept, $\Delta\sim 0.2- 0.3$, expected for the bare Pomeron of the resummed NLL$(1/x)$ BFKL approach \cite{BFKL}.  In comparison with one-pomeron exchange, non-enhanced eikonal absorption suppresses the growth of the amplitude with energy.  Therefore to describe the same data we need a larger intercept ($\Delta \simeq 0.11$).
 On the other hand, we already include the absorption caused by enhanced diagrams in our `effective' pomeron. As a result
 we expect a smaller effective intercept than that given by BFKL.  Similar arguments apply to the slope of the effective trajectory, leading to a value ($\alpha'_P \lapproxeq 0.1 ~\GeV^{-2}$)
 intermediate between the BFKL prediction($\alpha'_P \to 0$)
 and the old one-pole parametrization \cite{DL1} ($\alpha'_P=0.25~ \GeV^{-2}$).

Thus we demonstrate that, using the GW formalism with a single `effective' pomeron, it is possible to describe the diffractive cross sections and to reproduce the energy dependence of \sd, $d\sigma_{\rm el}/dt$, $\sigma_{\rm tot}$ and $d\sigma/d(\Delta\eta)$ in a large energy range from the CERN-ISR up to the LHC. The energy dependence of \se and $\sigma_{\rm tot}$ is controlled by the intercept and slope of the effective pomeron trajectory. The energy behaviour of low-mass dissociation is controlled by the properties of the GW eigenstates $\phi_i$ and the $\phi_i$-pomeron coupling $\gamma_i=(1\pm \gamma)$ or $k_i$ in Table \ref{tab:1}, while the energy dependence of high-mass dissociation is driven by the multi-pomeron effects specified by $\lambda$ and the gap survival factor $S^2$.

The evaluation of the gap survival factor $S^2$ is important in order to calculate the cross section of the various exclusive processes, like exclusive Higgs production or the recently. measured $WW$ production via $\gamma$ exchange \cite{CMSWW} or exclusive $J/\psi$ and $\Upsilon$ production \cite{exclJ}\footnote{In processes mediated by $\gamma$-exchange, the value of $S^2$ is not small, since the process is dominated by contributions from large distances in impact parameter space, $b$. Nevertheless the corrections are not negligible, and may be about 20-30$\%$.}. In turns out that the probability of gap survival is very sensitive to the detailed structure of the GW eigenstates. In the different versions of the model the value $S^2$ may vary by a factor of 3, while the description of the other observables is essentially the same. In order to improve the determination of the GW eigenstates, it is desirable to measure \sd and its $t$ dependence more precisely.

{\bf We conclude} that all 4 versions of the 2-channel eikonal presented here satisfactorily describe the available diffracive data, including the diffractive dip in the elastic scattering cross section and the energy dependence of the cross section for low-mass dissociation, \sd, within its present uncertainties.

Physically the coupling of the GW eigenstates to the `effective pomeron should be extended to allow for their energy dependence (since the effective pomeron is not a simple pole, but a more complicated object whose properties have an intrinsic energy dependence). For this reason model 4 is favoured.

 \section*{Appendix A: Observables in terms of GW eigenstates}
In the Good-Walker approach \cite{GW}, low-mass diffractive dissociation is described in terms of so-called diffractive (or GW) eigenstates, $|\phi_i\rangle$ with $i=1,n$, that diagonalize the $T$-matrix, and so only undergo `elastic' scattering. On the other hand high-mass dissociation is described in terms of multi-pomeron diagrams. We discuss our treatment of high-mass dissociation in Appendix B.

In this Appendix we recall the GW formalism.  First, the incoming `beam' proton wave function is written as a superposition of the diffractive eigenstates
\begin{equation}
|p\rangle~=~\sum a_i |\phi_i\rangle,
\label{eq:ai}
\end{equation}
and similarly for the incoming `target' proton. In this paper we use two diffractive eigenstates, $i=1,2$. In terms of this $2$-channel eikonal model, the $pp$ elastic cross section has the form
\be
\frac{d\sigma_{\rm el}}{dt}~=~\frac{1}{4\pi}  \left| \int d^2b~e^{i\q_t \cdot \b} \sum_{i,k}|a_i|^2 |a_k|^2~(1-e^{-\Omega_{ik}(b)/2}) \right|^2,
\label{eq:dsel}
\ee
where $-t=q_t^2$, and the opacity $\Omega_{ik}(b)$ corresponds to one-pomeron-exchange between states $\phi_i$ and $\phi_k$ written in the $b$-representation. Also we have
\be
\sigma_{\rm el}~=~  \int d^2b \left|~\sum_{i,k}|a_i|^2 |a_k|^2~(1-e^{-\Omega_{ik}(b)/2}) \right|^2,
\ee
\be
\sigma_{\rm tot}~=~  2\int d^2b~\sum_{i,k}|a_i|^2 |a_k|^2~(1-e^{-\Omega_{ik}(b)/2}) 
\label{eq:tot}
\ee
 and the `total' low-mass diffractive cross section
 \be
\sigma_{\rm el+SD+DD}~=~  \int d^2b ~\sum_{i,k}|a_i|^2 |a_k|^2 ~\left|(1-e^{-\Omega_{ik}(b)/2}) \right|^2,
\ee
where SD includes the single dissociation of both protons.
So the low-mass diffractive dissociation cross section is
\be
\sigma^{\rm D}_{{\rm low}M}~=~\sigma_{\rm el+SD+DD}-\sigma_{\rm el},
\ee
where $\sigma_{\rm el+SD+DD}$ corresponds to all
possible low-mass dissociation caused by the dispersion of the Good-Walker
eigenstate scattering amplitudes. As mentioned in footnote 2, this
corresponds to $M_{\rm diss}\sim 2 - 3$ GeV for the TOTEM data \cite{TOTEM}.

\section*{Appendix B: Formulae for high-mass dissociation}
The process $pp \to X+p$, where one proton dissociates into a system $X$ of 
{\it high-mass} $M$ is conventionally studied in terms of the triple-Pomeron coupling, shown as the dot between the dashed lines  in Fig. \ref{fig:3Rb}(a). 
In the absence of absorptive corrections, the
corresponding cross section is given by
\be
\frac{M^2 d\sigma}{dtdM^2}~=~g_{3P}(t)\beta(0)\beta^2(t)~\left(\frac{s}{M^2}\right)^{2\alpha(t)-2}~\left(\frac{M^2}{s_0}\right)^{\alpha(0)-1},
\label{eq:3P}
\ee
where $\beta(t)$ is the coupling of the Pomeron to the proton and $g_{3P}(t)$ is the triple-Pomeron coupling.
The coupling $g_{3P}$ is obtained from a fit to lower energy data.  Mainly it is the data on proton dissociation taken at the CERN-ISR with energies from $23.5 \to 62.5$ GeV.

The problem, in the above determination of $g_{3P}$, is that this is an effective vertex with coupling
\be
g_{\rm eff}~=~g_{3P}*S^2
\ee
which already includes the suppression $S^2$ -- the probability that no other secondaries, simultaneously produced in the same $pp$ interaction, populate the rapidity gap region.  Recall that the survival factor $S^2$ depends on the energy of the collider.  Since the opacity $\Omega$ increases with energy, the number of multiple interactions, $N \propto \Omega$, grows\footnote{This is because at larger optical density $\Omega$ we have a larger probability of interactions.}, leading to a smaller $S^2$.  Thus, we have to expect that the naive triple-Pomeron formula with the coupling \cite{abk,KKPT}, 
measured at relatively low collider energies will appreciably overestimate the cross section for high-mass dissociation at the LHC. A more precise analysis \cite{luna} accounts for the survival effect $S^2_{\rm eik}$ caused by the eikonal rescattering of the fast `beam' and `target' partons.  In this way, a coupling $g_{3P}$ about a factor of 3 larger than $g_{\rm eff}$ is obtained, namely $g_{3P} \simeq 0.2g_N$, where $g_N$ is the coupling of the Pomeron to the proton. The analysis of Ref. \cite{luna} enables us to better take account of the energy dependence of $S^2_{\rm eik}$.

\begin{figure} 
\begin{center}
\vspace*{-3cm}
\includegraphics[height=12cm]{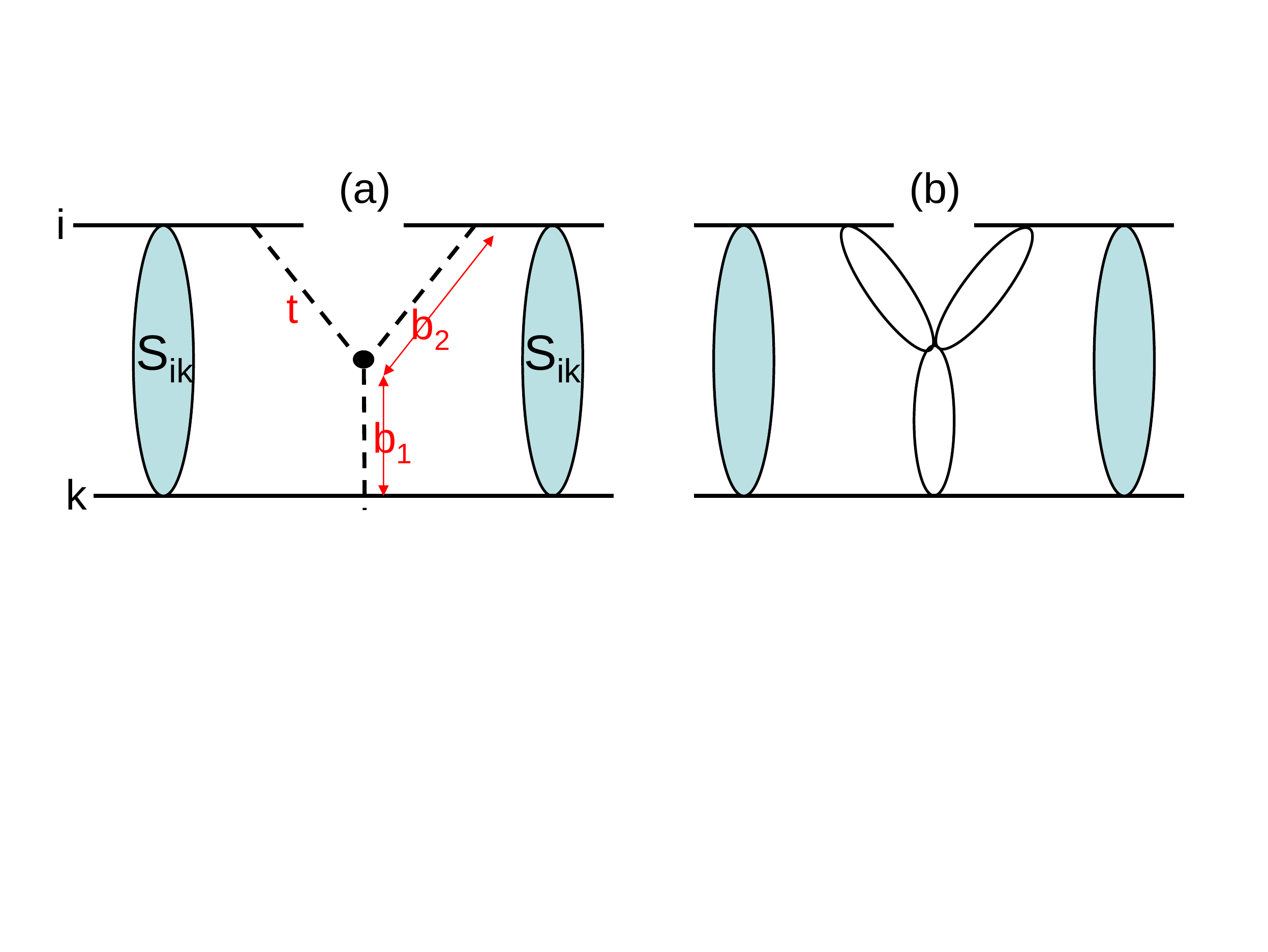}
\vspace*{-5.5cm}
\caption{\sf (a) A schematic diagram showing the notation of the impact parameters arising in the calculation of the screening corrections to the triple-pomeron contributions to the cross section; (b) a symbolic diagram of multi-pomeron effects. }
\label{fig:3Rb}
\end{center}
\end{figure}
To account for the absorptive effect, it is easier to work in the impact parameter, $b$, representation. 
To do this we follow the procedure of Ref. \cite{luna}. We first take Fourier transforms with respect to the impact parameters specified in Fig. \ref{fig:3Rb}(a). Then (\ref{eq:3P}) becomes 
\be
\frac{M^2 d\sigma_{ik}}{dtdM^2}~=~A\int\frac{d^2b_2}{2\pi}e^{i\vec{q}_t \cdot \vec{b}_2} F_i(b_2)\int\frac{d^2b_3}{2\pi}e^{i\vec{q}_t \cdot \vec{b}_3} F_i(b_3)\int\frac{d^2b_1}{2\pi} F_k(b_1),
\label{eq:3Rb}
\ee
where $F_i(b)$ is described by the opacity corresponding to the interaction of eigenstate $\phi_i$ with a intermediate parton placed at the position of the triple-pomeron vertex, while $F_k(b)$ describes the opacity of eigenstate $\phi_k$ from the proton which dissociates and interacts with the same intermediate parton. After integrating (\ref{eq:3Rb}) over $t$, the cross section becomes
\be
\frac{M^2 d\sigma_{ik}}{dM^2}~=~A\int\frac{d^2b_2}{\pi}\int\frac{d^2b_1}{2\pi} |F_i(b_2)|^2 F_k(b_1) \cdot S_{ik}^2(\vec{b}_2-\vec{b}_1),
\label{eq:result}
\ee
where here we have included the screening correction $S_{ik}^2$, which depends on the separation in impact parameter space, $(\vec{b}_2-\vec{b}_1)$, of states $\phi_i,\phi_k$ coming from the incoming protons
\be
S_{ik}^2(\vec{b}_2-\vec{b}_1)~\equiv~{\rm exp}(-\Omega_{ik}(\vec{b}_2-\vec{b}_1)).
\ee
If we now account for more complicated multi-pomeron vertices, coupling $m$ to $n$ pomerons, and assume an eikonal form of the vertex with coupling
\be
g^m_n=(g_N\lambda)^{m+n-2},
\label{eq:gmn}
\ee
then we have to replace $F_i$ by the eikonal elastic amplitude and $F_k$ by the inelastic interaction probability.  That is, instead of $F_i=\Omega_i(b_2)$ and $F_k=\Omega_k(b_1)$, we put
\be
F_i \to 2(1-e^{-\Omega_i(b_2)/2}), ~~~~~~~~~~ F_k \to (1-e^{-\Omega_k(b_1)}).
\ee
Fig. \ref{fig:3Rb}(b) symbolically indicates multi-pomeron couplings.
In (\ref{eq:gmn}), $g_N$ is the proton-pomeron coupling and $\lambda$ determines the strength of the triple-pomeron coupling.

\section*{Acknowledgements}

We thank Lucian Harland-Lang for discussions. MGR thanks the IPPP at the University of Durham for hospitality. This work was supported by the grant RFBR 11-02-00120-a
and by the Federal Program of the Russian State RSGSS-4801.2012.2.

%
\end{document}